\documentclass[10pt,journal]{IEEEtranTCOM}

\usepackage[english]{babel}
\usepackage[usenames]{color}
\usepackage[cp1250]{inputenc}
\usepackage{amsfonts}
\usepackage{amsthm}
\usepackage{graphicx}
\usepackage{epsfig}
\usepackage{mathrsfs}
\usepackage{amsmath}
\usepackage{algorithm}
\usepackage{algorithmic}
\usepackage{hyperref}

\theoremstyle{plain}

\begin{document}
\newcommand{\bea}{\begin{eqnarray}}
\newcommand{\eea}{\end{eqnarray}}
\newcommand{\be}{\begin{equation}}
\newcommand{\ee}{\end{equation}}
\newcommand{\beas}{\begin{eqnarray*}}
\newcommand{\eeas}{\end{eqnarray*}}
\newcommand{\bs}{\backslash}
\newcommand{\bc}{\begin{center}}
\newcommand{\ec}{\end{center}}
\def\SC {\mathscr{C}}

\title{Diffusion models for atomic scale electron currents in semiconductor, p-n junction}
\author{\IEEEauthorblockN{Jarek Duda}\\
\IEEEauthorblockA{Jagiellonian University,
Golebia 24, 31-007 Krakow, Poland,
Email: \emph{dudajar@gmail.com}}}
\maketitle

\begin{abstract}
While semiconductor electronics is at heart of modern world, and now uses 5nm or smaller processes of single atoms, it seems there are missing models of actual electron currents in these scales - which could help with more conscious design of future electronics. This article proposes such practical methodology allowing to model approximated electron flows in semiconductor, nonlinear Ohm law in p-n junction, and hopefully more complex systems e.g. built of transistors. It assumes electron hopping between atoms using Maximal Entropy Random Walk based diffusion - chosen accordingly to (Jaynes) maximal entropy principle, this way leading to the same stationary probability density as quantum models.
Due to Anderson-like localization in nonhomogeneous lattice of semiconductor, electrons are imprisoned in entopic wells, e.g. requiring to exceed a potential barrier for conductance.

\end{abstract}
\section{Introduction}
While historically electronics was focused on much larger scales and heuristic models~(\cite{t1,t2,t3}), modern processors use e.g. 5nm process with perspective of further reductions, what means scales of single atoms - suggesting to include also quantum effects into considerations. Literature search for such conductance models was unsuccessful, complete quantum modelling is extremely costly from computational perspective, hence there is proposed relatively inexpensive intermediate modelling framework, which includes quantum stationary probability distribution, which is usually localized (Anderson~\cite{anderson}).

\begin{figure}[t!]
    \centering
        \includegraphics{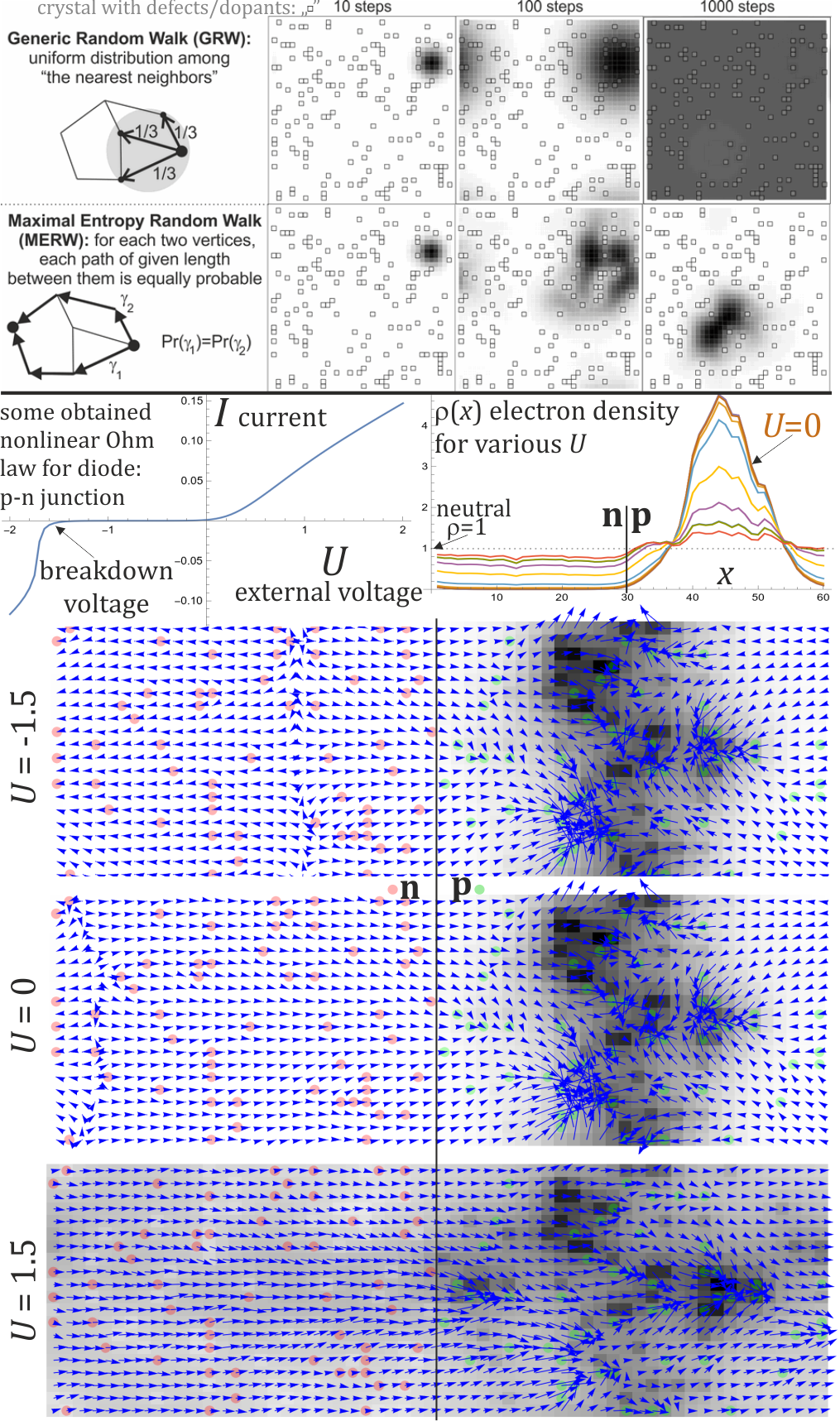}
        \caption{\textbf{Top}: two diffusion philosophies - standard GRW maximizing local entropy production (for each vertex) by using uniform edge ensemble, and MERW maximizing mean entropy production by using uniform path ensemble. Right: example of their evolution on defected $40\times 40$ lattice - all but the marked nodes have self-loop (degree 5 or 4 for marked), in analogy to dopant atoms in semiconductor. While GRW leads to nearly uniform stationary probability distribution, for MERW it is exactly as for quantum ground state, having Anderson-like localization imprisoning electrons in entopic wells - until applying threshold breakdown voltage. \textbf{Bottom}: example of obtained nonlinear current-potential dependence for MERW model on $60\times 30$ lattice. On the right there are electron densities - strongly localized for $\Delta V=0$, but equalizing this density for large $|\Delta V|$ allowing for conductance, much easier in $n\to p$ forward bias. Below grayness and arrows represent densities and local currents. }
       \label{merw}
\end{figure}

Specifically, we focus on Maximal Entropy Random Walk (MERW) type diffusion - chosen accordingly to the (Jaynes) maximal entropy principle. While for a regular lattice its predictions would be the same as for standard diffusion, referred here as Generic Random Walk (GRW), for defected lattice e.g. semiconductor with dopants as exchanged single atoms, predictions are very different as we can see in examples in Fig. \ref{merw}. While GRW leads to nearly uniform stationary probability density, MERW predicts stationary density as quantum mechanics - with strong Anderson-like localization property, confirmed experimentally for semiconductor with STM (scanning tunneling microscope)~\cite{exp}.

For defected lattice of semiconductor crystal, without such localization (GRW) tiny potential would already lead to electron conductance with (linear) Ohm law. In contrast, in reality this conductance is often blocked - here with Anderson-like localization, requiring some breakdown voltage to overcome the localization to allow for electron flow, getting very nonlinear voltage-current dependence.

This article extends on \cite{cond} MERW/GRW conductance simulator, among others by mean-field self-interaction (MFSI) inclusion of potential created by electron density, reflective boundary conditions allowing for flat boundary potentials in equilibrium, and finally application to practical systems like diode as p-n junction.

\section{Maximal Entropy Random Walk (MERW)}
\begin{figure}[t!]
    \centering
        \includegraphics{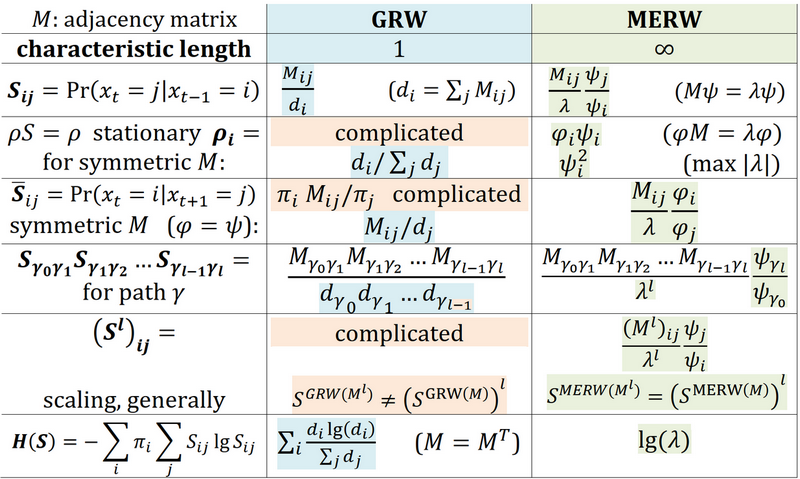}
        \caption{Gathered basic formulas for standard random walk/diffusion (GRW) and discussed MERW.        }
       \label{table}
\end{figure}

This section contains brief introduction to MERW~\cite{merwprl}, which now has many applications ($\approx 200$ citations), for example can be imagined as random walk along Ising sequence~\cite{ising}. The basic formulas (also standard GRW) are gathered in Fig. \ref{table}.

Deeper discussion can be found e.g. in \cite{myphd} - also for possible expansions of framework here, like continuous limit leading to Schr\"{o}dinger equation, or potentials varying in time for high frequency electronics.

Here we focus on single random walker. In the next section in mean-field way we will add self-interaction: treating stationary probability density of this walker also as charge density contributing to potential through Poisson equation - leading to more uniform electron density due to Coulomb repulsion.

\subsection{Maximal entropy principle and Boltzmann ensemble}
Let us start with basic question: "having $n$ white and black balls, what is the safest assumption for $p$ as percentage of white balls?". There are $2^n$ possibilities, splitting them into subsets of $pn$ white balls and assuming Stirling approximation ($n!\approx (n/e)^n$), we get approximate sizes of these subsets:
$$ {n\choose p} = \frac{n!}{(pn)!((1-p)n)!} \approx \exp(n h(p)) $$
for $h(p)=-p \ln(p)-(1-p)\ln(1-p)$ being Shannon entropy, which has single maximum: $h(1/2)=\ln(2)$. The $p=1/2$ subset grows with the highest exponent - asymptotically $(n\to\infty)$ completely dominating all the sequences, hence $p=1/2$ is the safest answer to our question.

This way we have derived a special case of Jaynes maximal entropy principle~\cite{jaynes}, saying that without additional knowledge, the safest choice of statistical parameters is the one maximizing entropy - as it describes  asymptotically dominating subset.

The above has assumed all e.g. $2^n$ possibilities are equally likely, while in physics we usually weight them with energy - maximizing entropy minus mean energy, which is $-\beta G$ for $G$ being Gibbs free energy:
\be \max_{(p_i):\sum_i p_i=1}\left(-\sum_i p_i\ln(p_i)-\sum_i p_i \beta E_i \right)=\ln\left(\sum_i e^{-\beta E_i} \right)\label{boltz}\ee
for $\ p_i\propto e^{-\beta E_i}$ Boltzmann ensemble, being at heart of statistical mechanics.

\subsection{Basic MERW as uniform path ensemble}
We would like to take the above to random walks, diffusion - which determine probability distribution among allowed paths.

Imagine a graph given by adjacency matrix $M_{ij}=1$ if there is edge from vertex $i$ to $j$, and 0 otherwise. Random walk as Markov process on this graph can be defined by stochastic matrix $S_{ij}=\Pr(\gamma_{t}=j |\gamma_{t-1} = i)$, which  satisfies $\sum_j S_{ij} =1$ for each vertex $i$. Such random walk usually leads to some stationary probability distribution $\rho$ being dominant eigenvector to eigenvalue 1: $\rho S=\rho$.

The question is how to choose this stochastic matrix $S$? A popular first guess is maximizing entropy locally for each vertex: $S_{ij}=M_{ij}/d_i$ for degree $d_i=\sum_j M_{ij}$. We refer to this choice as Generic Random Walk (GRW).

However, such local entropy maximization not necessarily maximizes mean entropy - averaged over vertex distribution $\rho$:
\be H=-\sum_i \rho_i \sum_j S_{ij} \ln(S_{ij}) \label{rwe}\qquad \textrm{entropy per step,}\ee
which maximization leads to discussed MERW stochastic matrix $S$. This entropy (per step) can be seen as of ensemble of paths $\gamma$ generated by $S$ stochastic process:
\be H=-\lim_{l\to \infty} \frac{1}{l} \sum_{\gamma=(\gamma_0..\gamma_l)} \Pr(\gamma) \ln(\Pr(\gamma)) \label{path} \ee
Entropy is maximized for uniform ensemble, this time of possible paths - hence MERW is equivalently uniform ensemble among paths on given graph.

\subsection{Used weighted MERW as Boltzmann path ensemble}
To add weights we go from uniform to Boltzmann ensemble - here of paths, exactly as for random walk along Ising sequence (useful e.g. to find its pattern distributions \cite{ising}).

Let us define $V_{ij}$ as energy of $i$-$j$ pair of vertices, e.g. $V_{ij}=(V_i +V_j)/2$ for point potential $V_i$. It allows to define energy of path $\gamma$ with discrete time $t\in\mathbb{Z}$ as $E(\gamma) = \sum_t V_{\gamma_{t-1} \gamma_t}$.

Entropy maximization (\ref{rwe}) now requires to subtract mean energy $\sum_{ij} \rho_i S_{ij} V_{ij}$. Inverting sign it becomes minimization of Gibbs free energy $G$ per step:
\be\min_S G\qquad\textrm{for}\qquad G
=\frac{1}{\beta}\sum_{ij} \rho_i S_{ij} (\ln(S_{ij})+\beta V_{ij}) \ee
Which is minimized for Boltzmann ensemble - now among paths. Its partition function $Z$ can be expressed with power of (analogous to minus tight-binding Hamilonian) popular in statistical mechanics~(e.g. \cite{baxter}) \be\textrm{\emph{transfer matrix}:}\qquad M_{ij}:=\exp(-\beta V_{ij}) \ee
$$Z_l:=\sum_{\gamma_0,\ldots,\gamma_l} \exp\left(-\beta(V_{\gamma_0 \gamma_1} + \ldots+ V_{\gamma_{l-1}\gamma_l})\right)=\sum_{\gamma_0,\gamma_l} \left(M^l\right)_{\gamma_0 \gamma_l}$$
Asymptotically the right hand side behaves $\sim \lambda^l$ for $\lambda$ being dominant eigenvalue of $M$. Finally equation (\ref{boltz}) allows to express free energy per step as:
\be\beta G =-\lim_{l\to \infty} \frac{\ln(Z_l)}{l} = -\ln(\lambda) \ee

\subsection{Sketch of combinatorial derivation of $S$, $\rho$ formulas}
\begin{figure}[t!]
    \centering
        \includegraphics{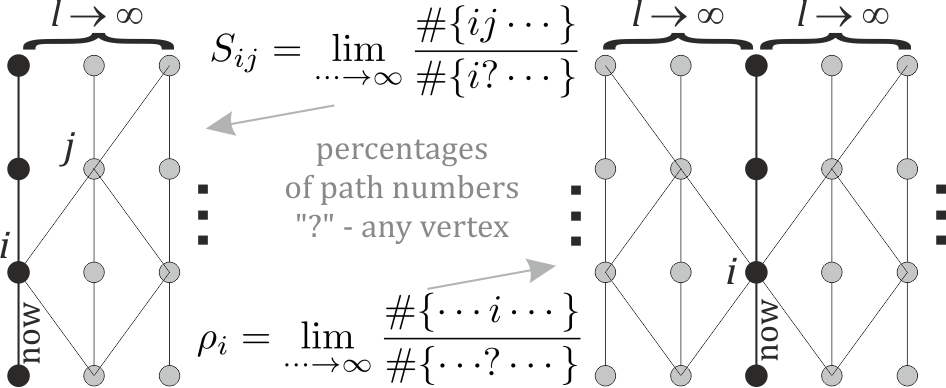}
        \caption{Sketch of derivation of $S_{ij}=\Pr(\gamma_{t}=j |\gamma_{t-1} = i)$ MERW stochastic matrix (left) and $\rho_i =\Pr(\gamma_t = i)$ stationary probability distribution (right) from statistics inside Boltzmann path ensemble in $l\to \infty$ limit.}
       \label{merwform}
\end{figure}

As in Fig. \ref{merwform}, let us present sketch of derivation of MERW as random walk given by Boltzmann ensemble of paths, exactly like for random walk along Ising-like sequence. Deeper discussion and expansions can be found e.g. in \cite{myphd}.


As mentioned, powers of transfer matrix $M_{ij}:=\exp(-\beta V_{ij})$ contain Boltzmann path ensemble:
\be (M^l)_{ij}=\sum_{\substack{\gamma_1\ldots \gamma_{l-1}\\ \gamma_0=i,\gamma_l=j}} \exp\left(-\beta \sum_{t=1}^l V_{\gamma_{t-1}\gamma_t}\right)\ee
Assuming connected aperiodic graph, Frobenius-Perron theorem says that $M$ has unique dominant left/right eigenvalues:
\be \max_{|\lambda|}:\qquad M \psi=\lambda\psi\qquad \phi^T M=\lambda \phi^T \ee
getting asymptotic $M^l \approx \lambda^l \phi \psi^T$, which allows to calculate $S_{ij}=\Pr(\gamma_{t}=j |\gamma_{t-1} = i)$ like in Fig. \ref{merwform} (for any $k$):
\be S_{ij}=\frac{\Pr(ij)}{\Pr(i)\ }=\lim_{l\to \infty} \frac{M_{ij} (M^l)_{jk}}{(M^{l+1})_{ik}}=\frac{M_{ij}}{\lambda} \frac{\psi_j}{\psi_i}\ee
and analogously stationary probability density from statistics inside such sequences (for any $j,k$):
\be \rho_i =\Pr(i)\propto \lim_{l\to\infty} (M^l)_{ji} (M^l)_{ik}\propto \phi_i \psi_i \ee
Figure \ref{table} gathers some formulas for standard random walk and MERW. The former seems more intuitive: walker assumes the same probability for all seen edges. In contrast to conscious walker, MERW should be imagine as statistical mechanics effective model: not knowing hidden complex walker behavior, the safest assumption is given be the maximal entropy principle.

\subsection{Sketch of optimization derivation of $S$, $\rho$ formulas}\label{opt}
Alternatively, we could treat this problem as maximization with 3 sets of constraints - for simplicity using $\beta=1$:
$$\min_{S,\rho}G(S,\rho)\qquad \textrm{for}\qquad G=\sum_{ij}\rho_i S_{ij}(\ln(S_{ij}) + V_{ij}) $$
$$\forall_i \sum_j S_{ij}=1\qquad \forall_j\, \rho_j=\sum_i \rho_i S_{ij}\qquad \sum_i \rho_i=1$$
Denoting the 3 types of Lagrange multipliers as $(\xi_i)$, $(\eta_j)$ and $\theta$, equating $S_{ij}$ derivative to 0 we get:
$$\ln(S_{ij})=\frac{\xi_i}{\rho_i}-V_{ij}-1+\eta_j$$
Substituting it to $\rho_i$ derivative equated to zero, we get:
$$\theta=\eta_i +\sum_j S_{ij}(\ln(S_{ij})+V_{ij}-\eta_j)=\eta_i +\sum_j S_{ij}\left(\frac{\xi_i}{\rho_i} - 1\right)$$
Using $\sum_j S_{ij}=1$ assumption we can simplify:
$$\theta=\eta_i + \xi_i/\rho_i -1\qquad\qquad \ln(S_{ij})=\theta -V_{ij}+\eta_j-\eta_i$$
Denoting $\psi_i=\exp(\eta_i)$, $\phi_i=\rho_i \exp(\xi_i/\rho_i)$, $\lambda=\exp(-\theta)$, $M_{ij}=\exp(-V_{ij})$, the constraints equations become left/right eigenequations, leading to formulas as before.

We have minimization here, but focus only on eigenvectors as exponents - enforcing positive coordinates, leading to dominant eigenectors as before.

\section{Defected lattice and self-interaction}
Let us now expand and apply discussed MERW methodology to electron conductance, \cite{cond} contains simplified simulator.
\subsection{Electron conductance on lattice e.g. crystal}
As the graph for our walker we would like to focus on lattices - with nodes representing single atoms, or maybe discretized larger systems. We work on 2D regular lattice (analogously in 3D) of size $n_x\times n_y$ hence containing $n=n_x n_y$ nodes.

The previously used vertex indexes $i,j$ correspond to such 2 (or higher) dimensional lattice vectors $i\equiv (x,y)$.

For simplicity we assume the walker can only jump to a nearest neighbor (4 in 2D, 6 in 3D) or stay in a given node: all but
$M_{(x,y),(x,y)},M_{(x,y),(x\pm 1,y)}, M_{(x,y),(x,y\pm 1)}$ terms of $n\times n$ matrix $M$ are zero - it is sparse matrix, of degree 5 in 2D, with cyclic boundary conditions especially in $x$ conductance direction to close the circuit.

Transfer matrix $M_{ij}=\exp(-\beta V_{ij})$ is nonzero only for these allowed jumps, the remaining can be imagined as corresponding to infinite energy $V_{ij}=\infty$.

The potential here has 3 types of contributions:
\be V =V^v + V^e + V^d \ee
\begin{itemize}
  \item $V^v$ describes potential characteristic for a given vertex e.g. as atoms of various types like dopants, ideally to be chosen based e.g. on orbital structures,
  \item $V^e$ corresponds to attached external voltage stimulating electron flow for conductance,
  \item $V^d$ describes potential resulting from electron density - while in literature it is usually chosen arbitrarily, here in the next subsection we include such self-interaction in electron density optimization.
\end{itemize}
The $V^v$ and $V^d$ contributions usually can be included as point-like e.g.: $V^v_{ij} =(V^v_i +V^v_j)/2$. We could also include direction dependence e.g. for Lorentz force in Hall effect.

In contrast $V^e$ rather requires special treatment. Naively it needs linear potential dependence like $V^e(x)=xU/n_x$ for attached external voltage $U$. However, it has problem when closing the circuit to allow circulation of electrons ($M$ needs cyclic boundary conditions in $x$) - requiring some analogue of e.g. battery. To avoid saw-like potential, we can subtract mean potential for each vertex, leaving constant gradient preferring jumps in one direction e.g.:
\be V^e_{(x,y),(x\pm 1,y)}=\exp(\mp \beta U/n_x) \ee
This term preferring jumps in one direction requires non-symmetric $M$, what corresponds to non-Hermitian Hamiltonian~\cite{non}.

\begin{figure}[t!]
    \centering
        \includegraphics{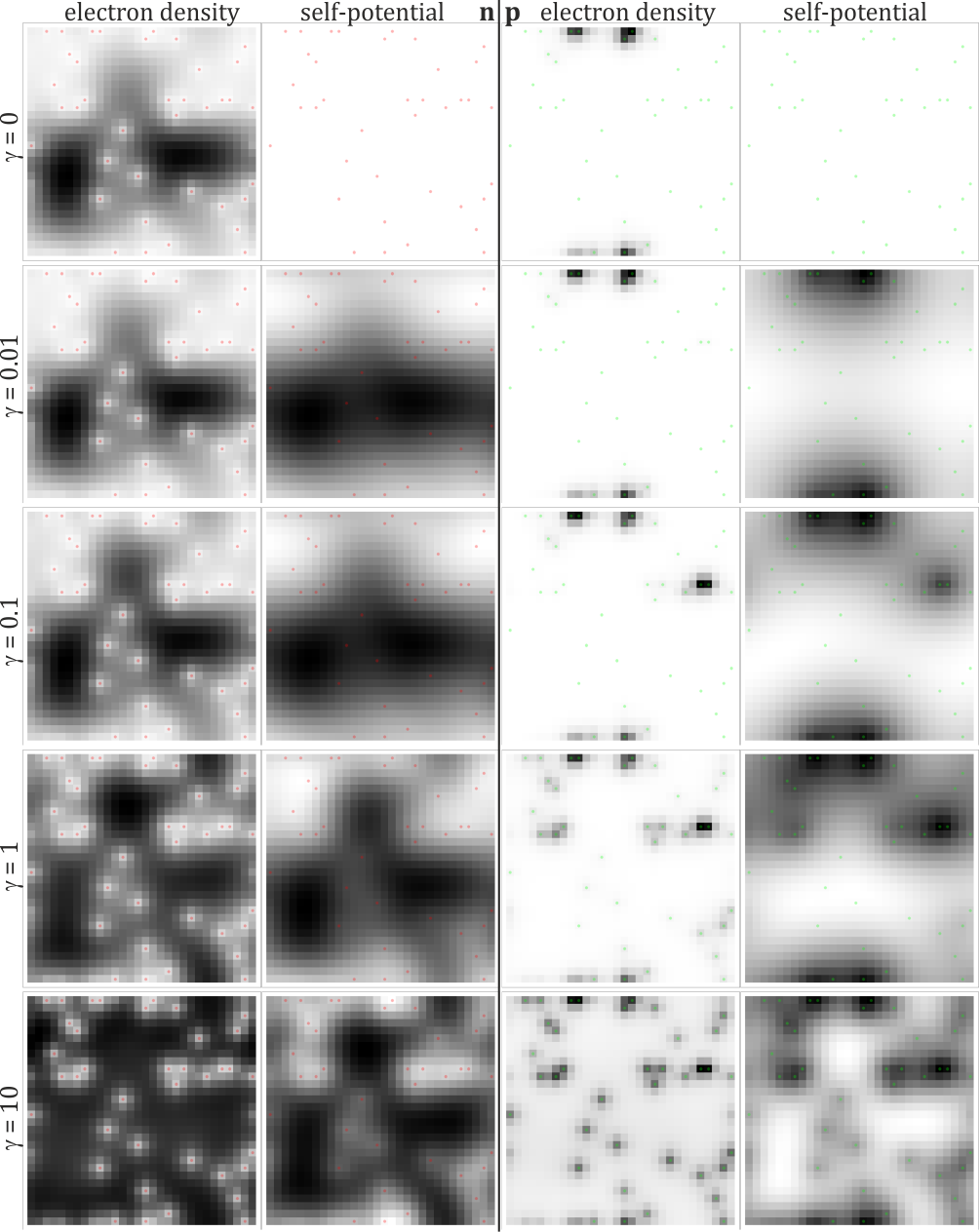}
        \caption{Visualized MERW electron densities $\rho$ and $V^d(\rho)$ potentials they produce (from Poisson equation) for $40\times 40$ lattice with cyclic boundary conditions and different strengths of self-interaction $\gamma$.
        All use the same randomly chosen defect pattern: n-type dopants (red dots of -0.5 potential, diagrams on left), and p-type dopants (green dots of +0.5 potential, diagrams on right).
        As expected, strong self-interaction $\gamma$ makes density more uniform  due to Coulomb repulsion. }
       \label{self}
\end{figure}

\begin{figure}[t!]
    \centering
        \includegraphics{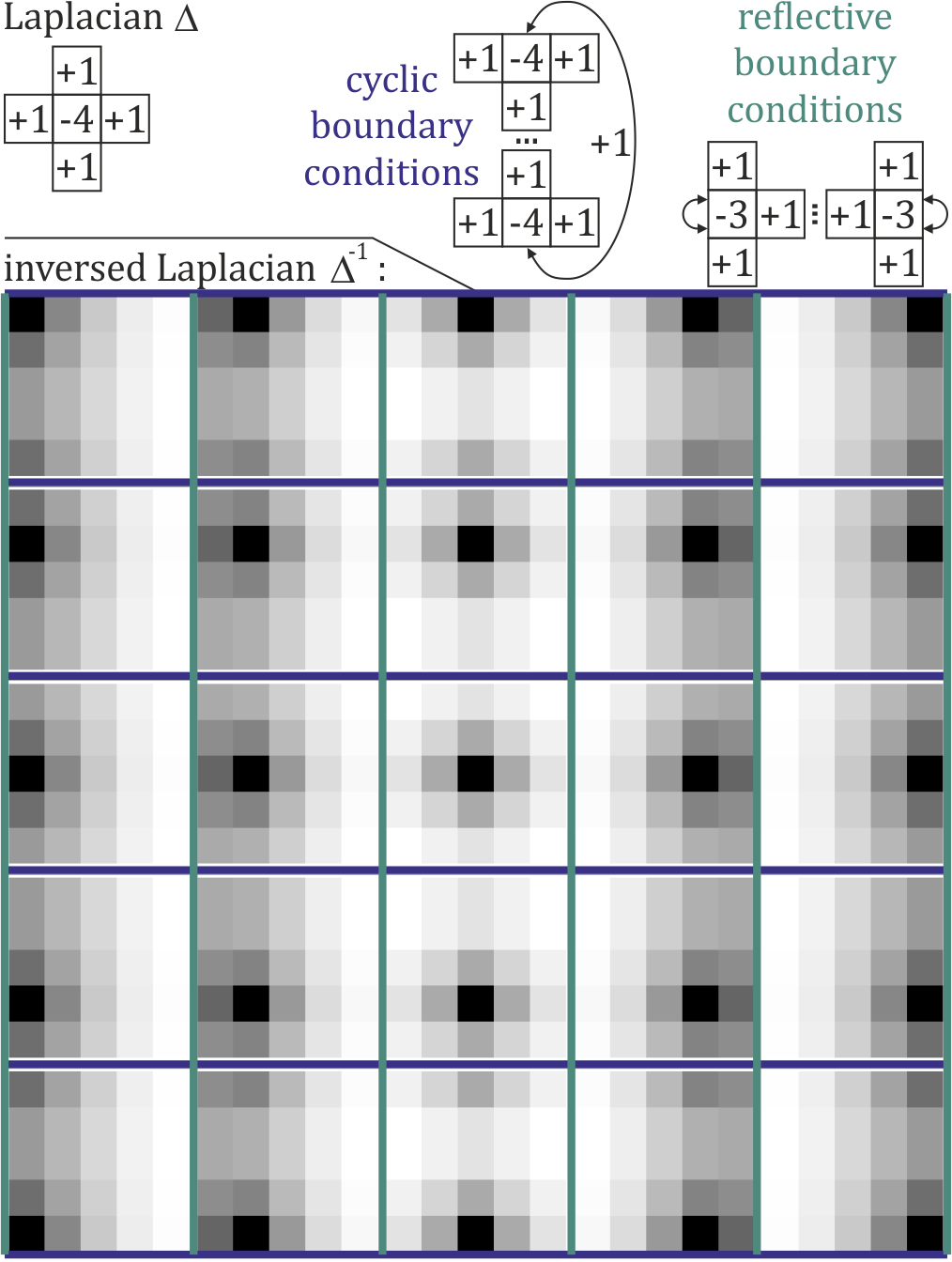}
        \caption{Top: discrete Laplacian used for $\Delta V \propto -\rho$ Poisson equation, and its used  modification for boundary conditions. In $y$ direction there is standard cyclic (for $M$ in both directions). To allow potential difference in $x$ conductance direction, for $\Delta$ there are used reflective boundary conditions - pretending reflected the same charges outside. Bottom: inverted Laplacian $\Delta^{-1}$ for $5\times 5$ lattice and such boundary conditions. To handle $\Delta$ being singular, what corresponds to freedom of choice of absolute potential, there is added 1 to all its terms before inverting.  }
       \label{ipois}
\end{figure}

\subsection{Adding mean-field self-interaction (MFSI)}
The discussed random walk has imagined single walker. However, in conductance we have multiple charged walkers: electrons, which Coulomb repulsion should make density more uniform, as presented in Fig. \ref{self}.

Hence we should interpret $\rho_i = \phi_i \psi_i$ stationary probability distribution also as representing charge density, bringing additional $V^d(\rho)$ contribution to potential, which can be calculated using Poisson equation: $\triangle V\propto - \rho$. Working on discrete lattice, we can disretize the Laplacian - e.g. in 2D use:
\be V^d_{x-1,y}+V^d_{x+1,y}+V^d_{x,y-1}+V^d_{x,y+1}-4 V^d_{x-1,y} =\gamma (\rho_{xy}-\rho^0_{xy}) \label{pois}\ee
for some constant $\gamma$, and $\rho^0$ charge density without the conductance electrons - for simplicity assumed as uniform $\rho^0_{xy}=1/n$.

These are $n=n_x n_y$ linear equation, we can find inverse of such discretized Laplacian as $n\times n$ matrix, finally getting:
\be V =V^0 + \gamma \Delta^{-1} \rho \quad\textrm{for}\quad V^0=V^v+V^e-\gamma \Delta^{-1} \rho^0 \ee
containing vertex potential describing node e.g. atom types $(V^v)$ and external voltage $(V^e)$.

Choosing boundary conditions is a subtle problem - while for $M$ at least in $x$ direction  cyclic are required, for Laplacian $\Delta$  there are used as in Fig. \ref{ipois}:
\begin{itemize}
\item Vertical ($y$): cyclic boundary conditions,
\item Horizontal ($x$): reflective boundary conditions treating edge as self-loop, this way pretending there are reflected charges behind it, what allows flat potential boundary behavior, different for both boundaries (for built-in voltage).
\end{itemize}
The $n\times n$ matrix $\Delta$ is singular here, what corresponds to freedom of choice of absolute potential. To invert it avoiding this issue, we can e.g. add 1 to all coordinates, what means adding 1 to the only 0 eigenvalue (to uniform eigenvector).\\

Numerical search of dominant left/right eigenvalue can be formulated as finding fixed point:
\be \rho_i=\phi_i \psi_i \qquad\quad \phi,\psi\textrm{ : eigenvectors of $M(\rho)$} \ee
including $V^d$ containing linear $\rho$ dependence.

A natural inexpensive approach is starting with some $\phi^0,\psi^0$ e.g. as uniform (or for assumed approximation of potential) and iterate $t\to t+1$ until some convergence condition:
\be \rho^t_i=\phi^t_i \psi^t_i \qquad \phi^{t+1},\psi^{t+1}\textrm{ : eigenvectors of }M^t=M(\rho^t) \ee
where e.g. dominant eigenvectors can be improved with single steps of power method like $\psi^{t+1}\propto M^t \psi^t$ and normalize. In practice it usually converges, however e.g. often oscillating, what suggests e.g. using some fractional steps, also in this moment there is no convergence guarantee. Generally, numerical approaches for MFSI require further work, alternative one by directly satisfying $2n+1$ constraints (2 eigenequations and normalization) is presented in Appendix.

There could also used some inexpensive approximations, like assuming parametrized family of potential only dependent on $x$.\\

\begin{figure}[t!]
    \centering
        \includegraphics{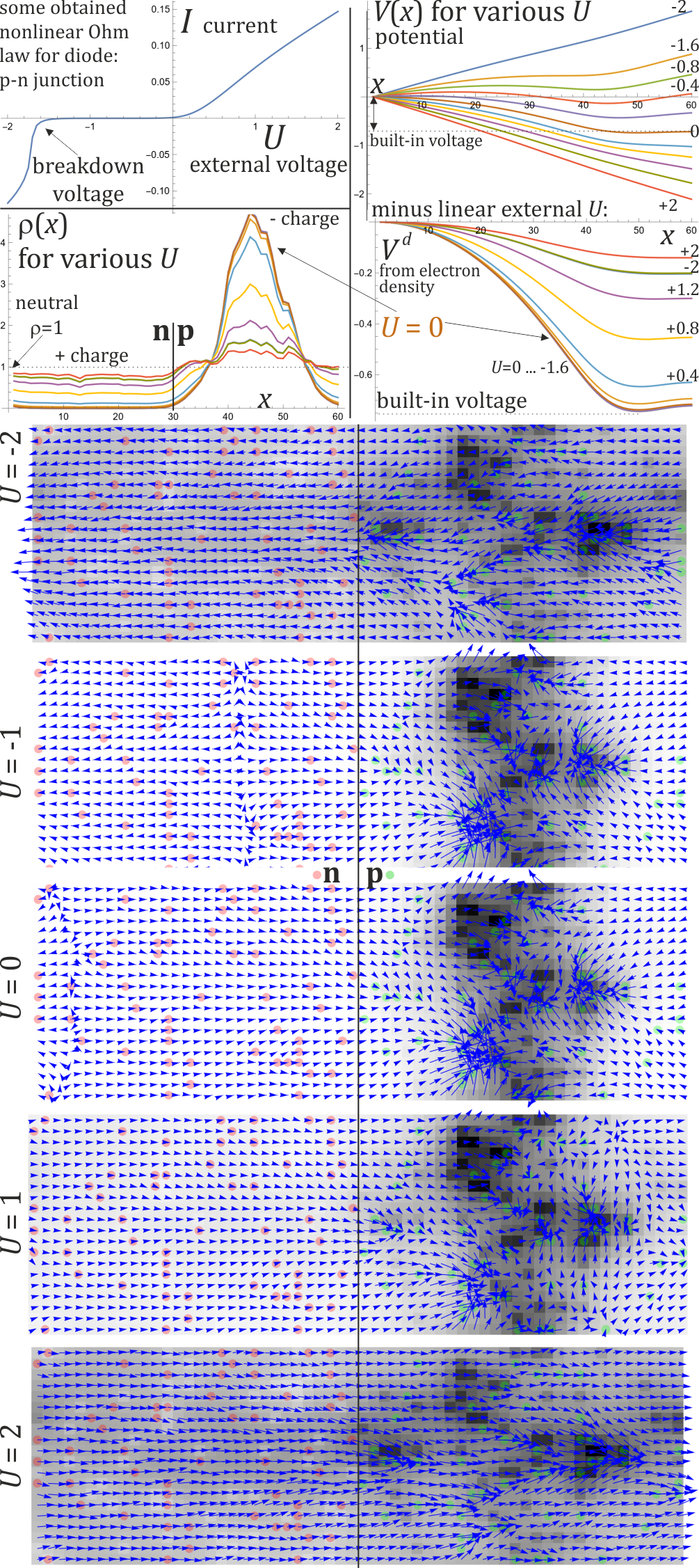}
        \caption{Example of p-n junction conductance dependance for $60\times 20$ lattice and various attached potential difference $\Delta V$. All use the same defect pattern: red dots denote +0.5 potential, green -0.5 ($\beta=10, \gamma=1$). Plotted arrows represent local electron currents (\ref{arrow}). We can see strong localization for low $\Delta V$ which prevents conductance, with equalizing density for larger $|\Delta V|$ (weakening built-in voltage), especially in $n\to p$ forward bias direction. }
       \label{cond}
\end{figure}

Fig. \ref{merw},\ref{cond}, \ref{reverse} present calculated example of p-n junction using such iteration until convergence, we can see voltage-current plot as expected for diode. Electrons are (Anderson) localized in p part in equilibrium, and their density becomes more uniform when attaching external voltage, weakening built-in voltage, more easily for forward bias.

The flow diagrams show fixed positions of p (green) and n (red) type dopants as atoms of different potential. Grayness presents electron densities. There are also arrows representing local currents as:
\be \rho_{xy}(S_{(xy),(x+1,y)}-S_{(xy),(x-1,y)},S_{(xy),(x,y+1)}-S_{(xy),(x,y-1)})
\label{arrow}
\ee
In reverse bias we can see kind of domain wall travelling with voltage - it has minimal density with symmetric diffusion in both directions, allowing for conductance when reaching junction.

\section{Conclusions and further work}
There was presented initial framework for modelling electron conductance e.g. in atomic scale, which is planned to be further developed, for example:
\begin{itemize}
   \item There was only presented simple diode example - should be also considered more complex e.g. transistor, trying to find applications of such models to improve technology e.g. by better control of defect distribution.
  \item Presented inexpensive framework is kind of between classical and quantum - recreates (localized) quantum stationary probability distribution, but neglects interference - there is required comparison with complete quantum models (like tight-biding) and experiments (e.g. temperature dependence, STM like~\cite{exp}), understanding inaccuracies and hopefully compensate them in inexpensive ways.
  \item The atomic potentials are very simple now, final models might require including atomic orbital structures, maybe multiple vertices per node corresponding to involved orbitals of given atom.
  \item There is assumed nearest-neighbors jumping between atoms, what seems reasonable for low voltage, but generally might require more complex modelling, up to more dense lattice and adding electron velocities like in Langevin model.
  \item There was discussed static modelling: of stabilized electron flows. In practice  electronics often works with very high frequencies. MERW for varying potential is discussed in \cite{myphd} and should be considered in the future.
  \item Better numerical methods are yet to be developed.
\end{itemize}

\appendix
To numerically include self-interaction, there was suggested cyclically updating potentials (Poisson equation) and dominant eigenvectors, hoping for some convergence. While it is inexpensive thanks to working on sparse matrices, there is no convergence guarantee. Hence here is proposed alternative numerical approach based on constraint satisfaction, however, it requires to work on dense matrices due to $\Delta^{-1}$ being dense. As in optimization derivation (\ref{opt}), to focus on dominant eigenvectors let us enforce them to have positive coordinates by directly optimizing their logarithms: $\Phi_i=\ln(\phi_i)$, $\Psi_i=\ln(\psi_i)$.
\subsection{Fixed potential}
For fixed $M_{ij}=\exp(-\beta V_{ij})$ we could just use standard numerical library to find dominant left/right eigenvectors $\phi$, $\psi$. To prepare for self-interaction, let us do it through iteration to satisfy the $n+1$ constraints $C=0$:
$$ C=\left((\lambda \psi_i -(M \psi)_i)_{i=1..n},\ 1-\sum_k \psi_k^2\right) $$
for $\psi_i =\exp(\Psi_i)$, the Jacobian $J$ made of derivatives is:
$$J_{i=1..n+1,\ j=1..n}:=\partial_{\Psi_j} C=((\lambda\delta_{ij}\psi_i - M_{ij} \psi_j)_{i=1..n},-2\psi_i^2)$$
$$J_{i=1..n+1,\ n+1}:=\partial_\lambda C =(\psi_{i=1..n},0)$$
Suggesting iteration (in practice with linear solve):
\be (\Psi,\lambda)\to (\Psi,\lambda)- J^{-1} C \label{step}\ee
Experimentally in 3-4 such steps $\|C\|$ usually gets from uniform $\psi$ down to $\sim 10^{-13}$ for random positive terms matrices of various sizes - for dense matrices allowing to approximate dominant eigenvectors of positive terms matrices a few times faster than standard Mathematica library with Arnoldi method.
\subsection{Plus self-interaction} We are now ready to add self-interaction: with no longer constant $M_{ij}=\exp(-\beta V_{ij})$: but this time depending on charge density $V=V(\rho)$ for
$$\rho_i=\phi_i \psi_i=\exp(\Phi_i+\Psi_i)\quad\textrm{normalized to}\quad\sum_i \rho_i =1$$

As discussed, self-interaction from Poisson equation, denoting $T=\gamma (\Delta)^{-1} /2$, can be written as:
\be V_{ij}(\rho) =V_{ij}^0 + \sum_k (T_{ik}+T_{jk})\rho_k \ee

For non-symmetric matrix here we need to use both eigenvectors, the constraint becomes $2n+1$ dimensional $C=0$:
\be C=\left((\lambda \phi -\phi^T M)_{i=1..n},\,(\lambda \psi -M \psi)_{i=1..n},\,1-\sum_k \phi_k \psi_k \right) \ee
We can analogously construct $2n+1\times 2n+1$ Jacobian from derivatives:
$$\partial_{\Phi_k} M_{ij}=\partial_{\Phi_k}\exp(-\beta V_{ij}) =-\beta M_{ij}(T_{ik}+T_{jk})\rho_k =\partial_{\Psi_k} M_{ij}$$
$$\partial_{\Phi_k} (\lambda \phi -\phi^T M)_i=\lambda \delta_{ik}\phi_i-\phi_k M_{ki}-\sum_j \phi_j \left(\partial_{\Phi_k} M_{ji}\right)$$
$$\partial_{\Psi_k} (\lambda \phi -\phi^T M)_i=-\sum_j \phi_j \left(\partial_{\Psi_k} M_{ji}\right) $$
$$\partial_{\Phi_k} (\lambda \psi -M \psi)_i=-\sum_j  \left(\partial_{\Phi_k} M_{ij}\right)\psi_j$$
$$\partial_{\Psi_k} (\lambda \psi -M \psi)_i=\lambda \delta_{ik}\psi_i- M_{ik}\psi_k-\sum_j  \left(\partial_{\Psi_k} M_{ij}\right)\psi_j$$
$$\partial_{\Phi_k}\left(1-\sum_i \phi_i \psi_i \right)=\partial_{\Psi_k} \left(1-\sum_i \phi_i \psi_i \right)=-\phi_k \psi_k $$
We can analogously perform step (\ref{step}), but this time it requires more caution e.g. changing step size until $\|C\|$ is indeed reduced, or use gradient method e.g. for $\|C\|^2$.

\begin{figure*}[t!]
    \centering
        \includegraphics[width=170mm]{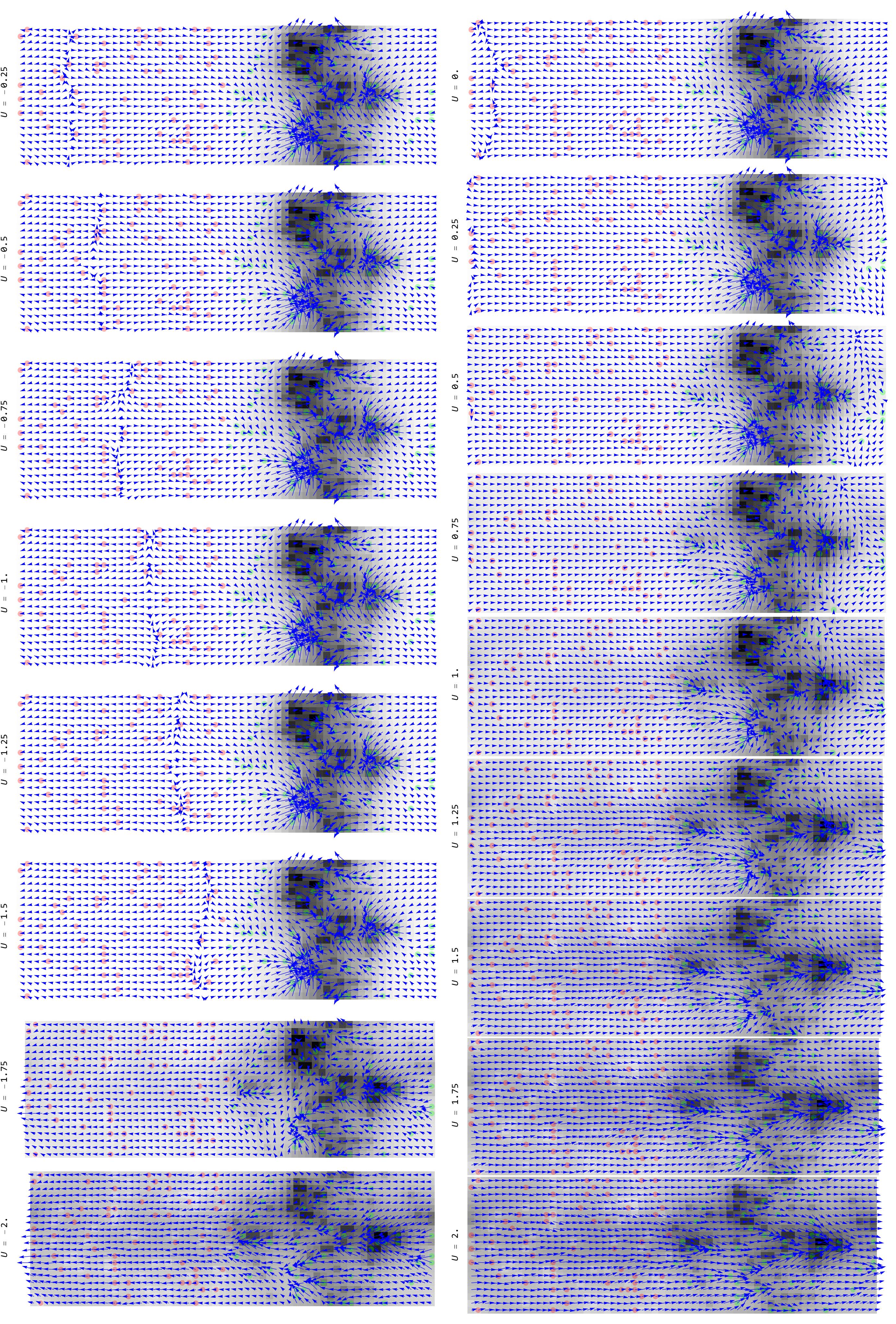}
        \caption{Higher voltage resolution of flows as in Fig. \ref{cond}, showing e.g. interesting domain wall shift before breakdown voltage in reverse bias. }
       \label{reverse}
\end{figure*}

\bibliographystyle{IEEEtran}
\bibliography{cites}
\end{document}